\documentclass[conference]{IEEEtran}
\IEEEoverridecommandlockouts
\usepackage{cite}
\usepackage{amsmath,amssymb,amsfonts}
\usepackage{bbm}
\usepackage{algorithmic}
\usepackage{graphicx}
\usepackage{textcomp}
\usepackage{xcolor}
\def\BibTeX{{\rm B\kern-.05em{\sc i\kern-.025em b}\kern-.08em
    T\kern-.1667em\lower.7ex\hbox{E}\kern-.125emX}}
\usepackage[caption=false, font=footnotesize]{subfig}
\usepackage{rotating}
\usepackage{pdflscape}
\usepackage{booktabs}
\usepackage{siunitx}
\usepackage{mathtools}
\usepackage{cuted}
\usepackage{multirow}
\usepackage{eqnarray}
\usepackage{amsmath}
\usepackage{placeins}
\usepackage{float}
\usepackage{lipsum}
\usepackage{flushend}
\usepackage{balance}
\usepackage{scalerel}
\usepackage{stfloats}
\usepackage{bigints}
\usepackage{textcomp}
\usepackage{tabularx}
\usepackage{cleveref}
\begin{document}

\title{Modeling the Doppler Shift in Cislunar Environment with\\  Gaussian Mixture Models
}

\author{
\IEEEauthorblockN{Baris Donmez}
\IEEEauthorblockA{\textit{Electrical Engineering Department} \\
\textit{Polytechnique Montréal}\\
Montreal, Canada \\
baris.donmez@polymtl.ca}
\and
\IEEEauthorblockN{Sébastien Loranger}
\IEEEauthorblockA{\textit{Electrical Engineering Department} \\
\textit{Polytechnique Montréal}\\
Montreal, Canada \\
sebastien.loranger@polymtl.ca}
\and
\IEEEauthorblockN{Gunes Karabulut Kurt}
\IEEEauthorblockA{\textit{Electrical Engineering Department} \\
\textit{Polytechnique Montréal}\\
Montreal, Canada \\
gunes.kurt@polymtl.ca}
}

\maketitle

\begin{abstract}
This study investigates the RF-based Doppler shift distribution characterization of the Lunar South Pole (LSP) based inter-satellite link (ISL) in varying inclination. Doppler shift in parts per million (ppm) is determined and analyzed, as it provides an independence from the carrier frequency. Due to unknown relative velocity states duration, the Gaussian Mixture Model (GMM) is found to be the best fitting distribution for ISLs with $1^\circ$ inclination interval Doppler shift with respect to a predetermined satellite. Goodness-of-fit is investigated and quantified with Kullback–Leibler (KL) divergence and weighted mean relative difference (WMRD) error metrics. Simulation results show that ISL Doppler shifts reach up to $\pm 1.89$ ppm as the inclination of the other orbit deviates higher from the reference orbit, inclining $80^\circ$. Regarding the error measurements of GMM fitting, the WMRD and KL divergence metrics for ISL take values up to 0.6575 and 2.2963, respectively.      
\end{abstract}

\begin{IEEEkeywords}
Doppler shift, Gaussian mixture model, inter-satellite links, lunar communication, radiofrequency (RF).
\end{IEEEkeywords}

\section{Introduction}
A strategically deployed constellation of lunar orbiters will play a critical role in future communication links. Unlike Earth-based orbits, lunar orbital dynamics differ significantly in both velocity and positional profiles, leading to distinct Doppler characteristics. These differences must be carefully characterized to ensure reliable communication performance.

The Lunar South Pole (LSP) is currently the most active region for lunar missions due to its scientific value, potential resource availability, and suitability for establishing long-term infrastructure \cite{donmez2025multiorbitercontinuouslunarbeaming}. As such, it has become a focal point for surveillance, exploration, and communication relay systems. Consequently, accurate Doppler characterization in this region is particularly important to support high data rate and low latency communication. Moreover, due to the orbital dynamics, the Doppler shift is time variant and follows a unique distribution even at the same altitude level. 

The low lunar orbit (LLO), which has an altitude of 100 km, is selected frequently for Cislunar communication systems as it offers relatively shorter distances than other orbit options \cite{orbittypes}. As radiofrequency (RF) technology has a bottleneck on path loss, opting for orbits with shorter path lengths is desirable~\cite{baris_hybrid}. 

In space communication systems, space vehicles onto which transmitters and/or receivers are placed move at very high speeds, such as kilometres per second. Hence, the impacts of the Doppler shifts must be considered in RF communication systems in space, as they significantly affect the decoding of the transmitted information.  

\begin{figure}[t!]
    \centering
    \includegraphics[width=0.90\linewidth]{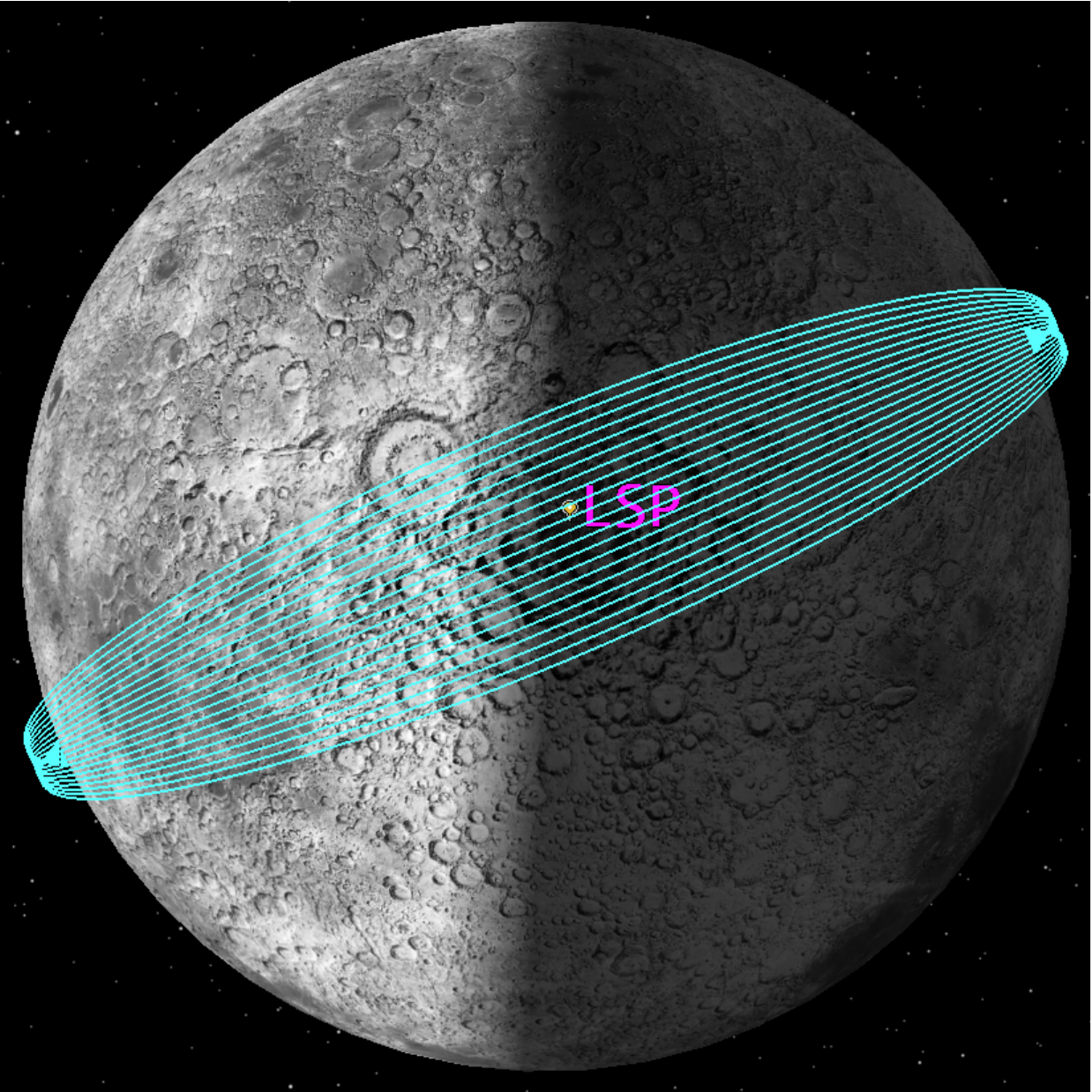}
    \caption{ Illustration of the proposed Cislunar environment showing all 21 orbits over which satellites move and the ground station located at LSP.}
    \label{fig:sysmod}
\end{figure}

\subsection{Literature Review}
It has been shown that the Doppler shift distribution is symmetrical around $0$ in many studies for satellite to ground \cite{al2024doppler}. Recently, channel modeling in \cite{mcbain2025stochastic} shows the analytical cumulative distribution function (CDF) and probability density function (PDF) of Doppler shift for non-homogeneous binomial point process, simplified general perturbations 4 and circular orbits in low Earth orbit (LEO) mega constellations. \cite{tanash2025statistical} showcases a tighter CDF bound of the Doppler shift magnitude given in \cite{khan2020stochastic}. 

On the other hand, the inter-satellite link (ISL) possesses a different Doppler characterization due to its unique relative motion. Due to the additional mobility, ISL Doppler shift characterization is investigated for free space optics (FSO) with Gamma Gamma distributions \cite{fernandes2023digitally}. 

\subsection{Motivation and Contributions}
Designing a realistic Cislunar system is challenging since astrodynamics requires numerous force parameters to be taken into account. Thus, we use the System Tool Kit (STK) \cite{ansys_stk} in which many external forces are considered to determine precise Doppler shift in both inter-satellite and satellite-to-ground station (GS) scenarios.  

To the best of our knowledge, the stochastic behaviour of the RF Doppler shift in the Cislunar environment will be analyzed first in our paper.  

Our contributions are as follows:
\begin{itemize}

\item A maximum likelihood estimator (MLE) for Doppler shift data fitting distribution parameters is developed for both single ISL and generalized ISL constellations with $N+1$ satellites.

\item In our Cislunar analyses, we utilize the STK high-precision orbit propagator (HPOP) for obtaining realistic ephemeris data. We incorporate the effects of external gravitational forces (i.e., the Earth), solar radiation pressure (SRP), and Lunar radiation pressure (i.e., thermal and albedo). The impacts of space debris are disregarded.

\item Doppler shifts in ppm, which are independent of the transmitting frequency, for 20 different ISL over the Moon are demonstrated in Fig. \ref{fig:2}. As the orbital inclination difference between the reference LLO orbiter (inclination:$80^\circ$) and another LLO orbiter increases, the Doppler shift increases too.      

\item The goodness-of-fit of the estimated GMM distributions with $K=5$ components is evaluated using Kullback–Leibler (KL) divergence and the weighted mean relative difference (WMRD) error metrics. The KL divergence error measurements range between 0.0111 and 2.2963, whereas WMRD errors are computed between 0.5794 and 0.6906 when the collection of all Doppler shifts exhibited in Fig. \ref{fig:3} is also included.

\item Satellite-to-ground station (GS) Doppler shift analysis shows that Doppler shifts are compressed at minimum \\-5.4505 ppm and maximum 5.4509 ppm values as demonstrated in Fig. \ref{subfig:3b}.  
 
\end{itemize}

\subsection{Paper Organization}
The remainder of our paper is organized as follows. In Section II, our proposed Cislunar scenarios are explained along with the proposed stochastic fitting model and corresponding error quantification model. In Section III, the distributions of Doppler shifts in inter-satellite scenarios are presented, along with the corresponding fitting curves. Furthermore, the goodness-of-fit of the proposed stochastic models is evaluated numerically. Finally, our study is concluded in Section IV.  

\begin{figure}[t!]
    \centering
    \includegraphics[width=0.90\linewidth]{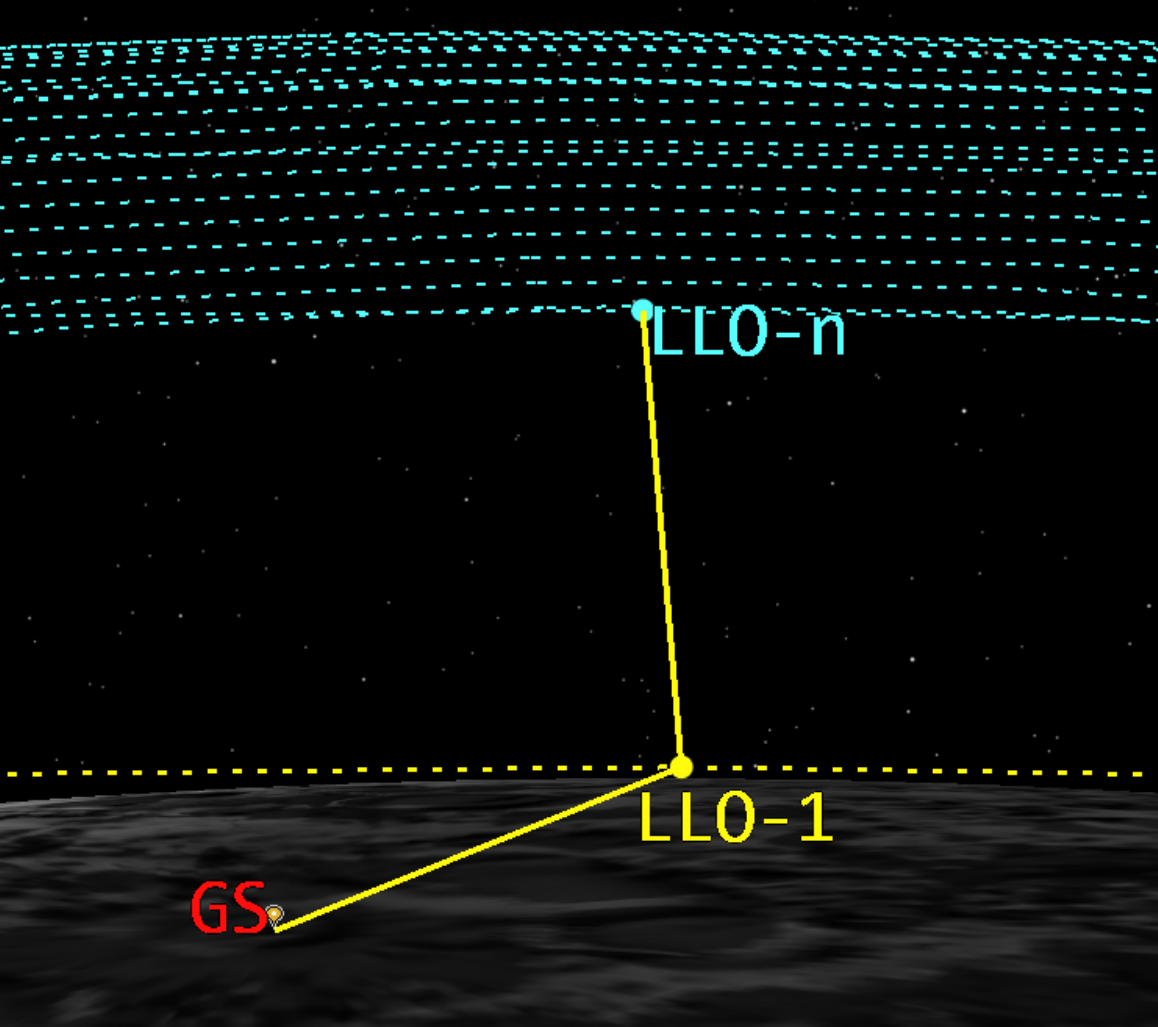}
    \caption{ Illustration of an ISL between $\textrm{LLO-1}$ and $\textrm{LLO-n}$, and the $\textrm{LLO-1}$-GS link.}
    \label{fig:sysmod2}
\end{figure}

\section{System Model}
$N+1$ LLO satellites orbiting around the Moon with $n\in N$ ISL aims at the connectivity with a predefined LLO satellite, namely $\textrm{LLO-1}$. Due to the mobility, $n$-th ISL is subject to a Doppler shift $f_{n,d} = \frac{\nu}{c} f_c$ where $\nu$ is the relative velocity, $f_c$ is the carrier frequency, and $c$ is the speed of light. A positive $f_{n,d}$ exists when satellites are moving toward $\textrm{LLO-1}$ during the approaching time interval $t_0 < t < t_1$, and becomes negative when they are moving away from $\textrm{LLO-1}$ during the subsequent interval $t_1 < t < t_2$ \cite{guven2023multi}.  

 The Doppler shift is dependent on the carrier frequency; thus, analyses based on parts per million (ppm) Doppler shift provide an independence from $f_c$ as follows \cite{doroshkin2019experimental}
\begin{equation}
    [\textnormal{ppm}] = \frac{f_{n,d}}{f_c} \times 1,000,000 
\end{equation}

The Gaussian mixture model (GMM) of the Doppler shift $f_{d}$ data $X=\{x_m\}$, $m=1,...,M$ is denoted as follows
\begin{equation}
    \hat{p}(X|\theta) = \sum_{k=1}^K \pi_{k} \mathbb{N}\left( x \, | \, \mu_{k} , \sigma_{k}^2 \right) \hspace{2mm} , 
\end{equation}
where $\pi_k$ is the mixture coefficient representing the probability of the $k$-th component, where $0< \pi_k <1 $ and $\sum_k \pi_{k} = 1$.  Parameters of GMM is denoted with $\mathbf{\theta}=\{ \mu_{k} , \sigma_{k}^2, \pi_{k}  \}$ subject to estimation. With the assumption that $M$ amount of transmitted package is exposed to the Doppler shift for each ISL, an ML estimator for $f_d$ is as follows
\begin{align} 
    \hat{p}(X\, | \, \mathbf{\theta}) = \prod_{m=1}^{M}  \sum_{k=1}^K \pi_{k} \frac{1}{\sqrt{2 \pi \sigma_{k}^2 }} \exp{\left( - \frac{\left(x_m - \mu_{k}\right)^2}{2 \sigma_{k}^2} \right)}  \hspace{2mm},  
\end{align}
where the log-likelihood is as follows 
{\small
\begin{align}
    \mathbb{L} &= \ln{\left(\hat{p}(X\, | \, \mathbf{\theta}\right))} \nonumber \\ 
   &= \sum_{m=1}^M \ln{\left[ \sum_{k=1}^K \pi_{k} \frac{1}{\sqrt{2 \pi \sigma_k^2 }} \exp{\left( - \frac{\left(x_m - \mu_{k}\right)^2}{2 \sigma_{k}^2} \right)}  \right]} .
    \label{loglike}
\end{align}
}
The partial derivative of Equation \ref{loglike} with respect to the first parameter of $\mu_k$ is the following

\begin{align}
\frac{\partial \mathbb{L}}{\partial \mu_{k}} = \sum_{m=1}^M \underbrace{\frac{\pi_{k} \mathbb{N}\left( x_m | \mu_{k} , \sigma_{k} \right) }{\sum_{j=1}^K \pi_{j} \mathbb{N}\left( x_m | \mu_{j} , \sigma_{j} \right)}}_{\rho_{mk}} \frac{x_m - \mu_{k}}{\sigma_{k}^2} \hspace{2mm}.
\label{parder}
\end{align}

Due to the missing prior information on relative velocity states between the $n$-th satellite and $\textrm{LLO-1}$, an iterative approach is required to estimate the posterior in Equation $\ref{parder}$.  

\begin{table}[!t]
  \sisetup{group-minimum-digits = 4}
  \centering
  \caption{Simulation Parameters}
  \label{tab:Kepler}
  \begin{tabular}{lllS[table-format=5]ll} 
    \toprule
    \toprule
     Semi-major axis & 1837.4 km  \\
     Eccentricity & $\approx0$   \\
     Argument of perigee & $0^{\circ}$  \\
          Longitude of ascending node 
 & $90^{\circ}$  \\
     True anomaly & $0^{\circ}$ \\
    Inclination ($\textrm{LLO-1}$ to $\textrm{LLO-21}$) & $80^{\circ}:1^{\circ}:100^{\circ}$ \\ 
    \midrule 
    \midrule 
    Simulation duration & 1 day  \\
    Sampling interval & 10 seconds  \\
   \midrule 
    Carrier frequency & 20 GHz  \\
    \bottomrule
    \bottomrule
\end{tabular}
\vspace{-0.2 cm}
\end{table} 

Expectation-Maximization (EM) algorithm is initialized with $k$-means algorithm by assuming the cluster number equals to the relative velocity state numbers with $    \hat{\mathbf{\theta}}_0 = \{ \hat{\mu}_{k} \, , \, \hat{\sigma}_{k}^2 \, , \, \hat{\pi}_{k} \}$ such that the first centroid selection (e.g. $\mu_{k=1}^{(0)}$) is a random point from $f_{d}$ data $X$. The next centroid selection probability is assigned following the distance rule
\begin{equation}
 \mu_k^{(0)} \leftarrow \mathbb{P}\mathbbm{r} (x) = \left( \frac{d(x)^2}{\sum^M d(x)^2} \right)  \hspace{1mm},
\end{equation}
where $d(x)$ is the Mahalanobis distance from a random sample to the closest centroid point. With the acquired $\mu_k^{(0)}$, initialize remaining parameters are following
\begin{equation}
 \sigma_{k}^{(0)} = \frac{1}{M_k} \sum_m^M (x_m - \mu_{k})^2 \hspace{1mm},\hspace{1mm} \pi_{k}^{(0)} = \frac{M_k}{M} \hspace{0.25cm},
\end{equation}
Following the initialization, the maximization step follows with the posterior information until the maximum iteration number
\begin{align}
\mu_{k}^{(j)} =  \frac{1}{M_k} \sum^M \rho_{mk} x_m \hspace{1.5cm} \nonumber \\
\sigma_{k}^{(j)} = \frac{1}{M_k} \sum^M \rho_{mk} (x_m - \mu_{k})^2 \hspace{0.5cm},\\ 
 \pi_{k}^{(j)} = \frac{M_k}{M} \hspace{3.3cm} \nonumber
\end{align}
where $M_k = \sum_{m=1}^M \rho_{mk}$ is the sample amount of $k$-th cluster which corresponds to the collected sample from the $k$-th relative velocity state in the $\textrm{LLO-1}$.

In a similar manner, a more generalized Doppler shift distribution is as follows 
\begin{equation}
    p(X_T|\theta_T) = [p(X_1|\theta_1), \cdots, p(X_N|\theta_N)] \hspace{2mm}.
\end{equation}
Estimating the parameters of $\hat{p}(X_T|\theta_T)$ follows the same procedure as before, using the maximum likelihood (ML) estimation function.

\subsection{Error Metrics}
Evaluation of the goodness-of-fit is performed by weighted mean relative difference \cite{kumar2004data} and Kullback–Leibler divergence~\cite{hershey2007approximating}.

Accordingly, ground truth PDF of  $p(X|\theta)$ and fitted curve PDF of $\hat{p}(X|\theta)$ is compared with the following metrics

\begin{align}
\textrm{WMRD}&=\frac{\sum{\left| {\hat{p}(X|\,\theta)}-{{p}(X|\,\theta)} \right|}}{\left( {1}/{2} \right)\times \sum{\left( {\hat{p}(X|\,\theta)}+{{p}(X|\,\theta)} \right)}} \hspace{2mm}, \\ \notag \\
\textrm{$D_{KL}$} &= \sum \hat{p}(X|\,\theta) \log{\frac{\hat{p}(X|\,\theta)}{{p}(X|\,\theta)}} \hspace{2mm}.
\label{KL}
\end{align}

\begin{figure}[!t]
\centering
\subfloat[]{
	\label{subfig:3a}
	\includegraphics[width=.22\textwidth]{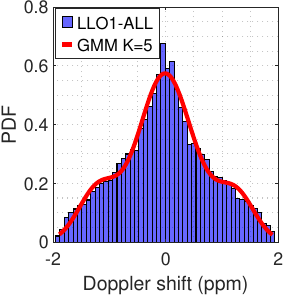}
	 }
\subfloat[]{
	\label{subfig:3b}
	\includegraphics[width=0.22\textwidth]{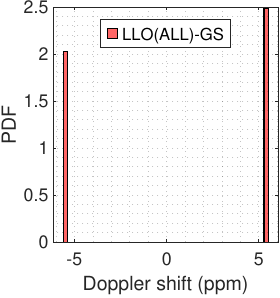}
	 }
\caption{(a) Distribution of all Doppler shifts in ppm between $\textrm{LLO-1}$ (incl:$80^{\circ}$) and each of the $N=20$ satellites with increasing inclinations.(b) Distribution of all Doppler shifts in ppm between $N+1=21$ satellites and the GS at LSP.}
\label{fig:3}
\end{figure}

\begin{figure*}[t!]
\centering
\includegraphics[width=1.0\textwidth]{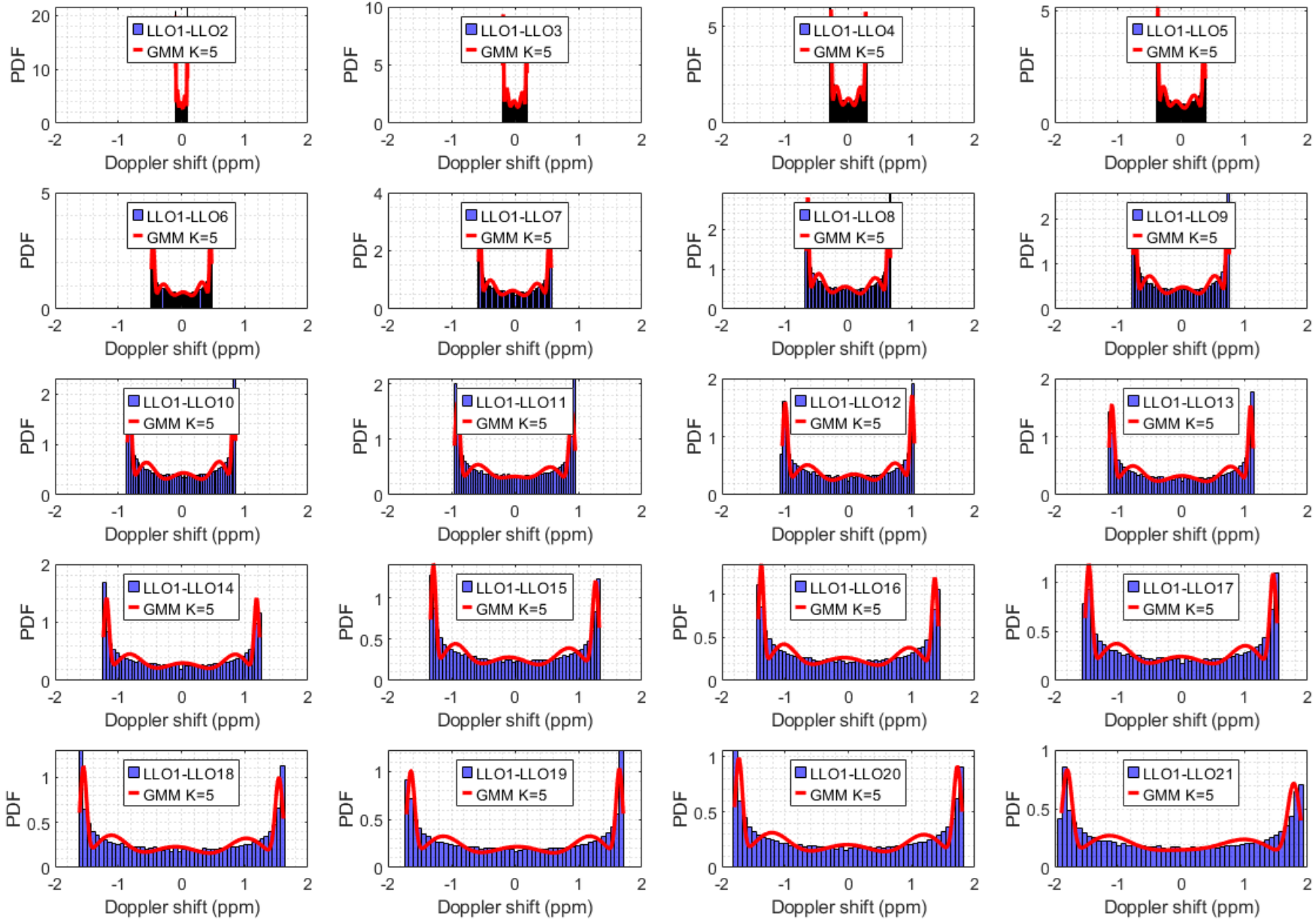}
\caption{Inter-satellite Doppler shifts in ppm between the reference LLO satellite of $\textrm{LLO-1}$ and other $N=20$ satellites with increasing inclinations. GMM with $K=5$  fits over the realistic data obtained from STK simulations.}
\label{fig:2}
\vspace{-0.35cm}
\end{figure*}

\begin{table}[h!]
\centering
\caption{ Error Quantification Metrics for PDF Estimation of ISLs}
\label{tab:my-Error}
\begin{tabular}{|l|ll|}
\toprule \toprule
\multicolumn{1}{|c|}{\multirow{2}{*}{\begin{tabular}[c]{@{}c@{}}\textbf{ISL}\end{tabular}}} &
  \multicolumn{2}{c|}{\begin{tabular}[c]{@{}c@{}}\textbf{Error Metrics}\end{tabular}} \\ \cline{2-3} 
\multicolumn{1}{|c|}{} &
  \multicolumn{1}{c|}{\textbf{WMRD}} &
  \multicolumn{1}{c|}{\textbf{KL Divergence}} \\ \midrule
LLO-1 and -2  & \multicolumn{1}{c|}{0.6575} & \multicolumn{1}{c|}{0.7057} \\ \midrule
LLO-1 and -3  & \multicolumn{1}{c|}{0.5812} & \multicolumn{1}{c|}{2.2963} \\ \midrule
LLO-1 and -4  & \multicolumn{1}{c|}{0.6093} & \multicolumn{1}{c|}{1.5421} \\ \midrule
LLO-1 and -5  & \multicolumn{1}{c|}{0.5996} & \multicolumn{1}{c|}{0.9872}  \\ \midrule
LLO-1 and -6  & \multicolumn{1}{c|}{0.6343} & \multicolumn{1}{c|}{0.4522} \\ \midrule
LLO-1 and -7  & \multicolumn{1}{c|}{0.6246} & \multicolumn{1}{c|}{0.5564} \\ \midrule
LLO-1 and -8  & \multicolumn{1}{c|}{0.6156} & \multicolumn{1}{c|}{0.5041} \\ \midrule
LLO-1 and -9  & \multicolumn{1}{c|}{0.6140} & \multicolumn{1}{c|}{0.3951} \\ \midrule
LLO-1 and -10 & \multicolumn{1}{c|}{0.6289} & \multicolumn{1}{c|}{0.3362} \\ \midrule
LLO-1 and -11 & \multicolumn{1}{c|}{0.6438} & \multicolumn{1}{c|}{0.1852} \\ \midrule
LLO-1 and -12 & \multicolumn{1}{c|}{0.6183} & \multicolumn{1}{c|}{0.2102} \\ \midrule
LLO-1 and -13 & \multicolumn{1}{c|}{0.6299} & \multicolumn{1}{c|}{0.1503} \\ \midrule
LLO-1 and -14 & \multicolumn{1}{c|}{0.6241} & \multicolumn{1}{c|}{0.1819}\\ \midrule
LLO-1 and -15 & \multicolumn{1}{c|}{0.6050} & \multicolumn{1}{c|}{0.2970} \\ \midrule
LLO-1 and -16 & \multicolumn{1}{c|}{0.5983} & \multicolumn{1}{c|}{0.3429} \\ \midrule
LLO-1 and -17 & \multicolumn{1}{c|}{0.5908} & \multicolumn{1}{c|}{0.2828} \\ \midrule
LLO-1 and -18 & \multicolumn{1}{c|}{0.6463} & \multicolumn{1}{c|}{0.1094} \\ \midrule
LLO-1 and -19 & \multicolumn{1}{c|}{0.6339} & \multicolumn{1}{c|}{0.1526} \\ \midrule
LLO-1 and -20 & \multicolumn{1}{c|}{0.6218} & \multicolumn{1}{c|}{0.2126} \\ \midrule
LLO-1 and -21 & \multicolumn{1}{c|}{0.5794} & \multicolumn{1}{c|}{0.2309} \\ \midrule 
ALL     & \multicolumn{1}{c|}{0.6906} & \multicolumn{1}{c|}{0.0111} \\ \bottomrule \bottomrule
\end{tabular}%
\vspace{-0.35cm}
\end{table}

\section{Results and Discussion}
The motions of the orbiters are simulated realistically by utilizing STK, and the simulation parameters are exhibited in Table \ref{tab:Kepler}. The inclination of the orbits ranges between $80^{\circ}$ and $100^{\circ}$ with a step size of $1^{\circ}$. The simulation duration is one day, during which an LLO satellite revolves $12$ times around the Moon, as its period is approximately $2$ hours \cite{orbittypes}.

Inter-satellite Doppler shift analyses are conducted between $\textrm{LLO-1}$ with inclination $80^ {\circ}$ and each of the remaining twenty LLO satellites, such as $\textrm{LLO-21}$ with inclination $100^{\circ}$. The Doppler shift distributions for each ISL are presented in Fig. \ref{fig:2} with a GMM fitting curve with $K=5$  components. Furthermore, the collection of all inter-satellite Doppler shifts and the GMM fitting curve with $K=5$ relative velocity state is exhibited in Fig. \ref{fig:3}. The Doppler shift reaches up to $\pm 0.0953$ ppm for the nearest ISL, which links $\textrm{LLO-1}$ and $\textrm{LLO-2}$ (inclination:$81^\circ$). On the other hand, computed Doppler shifts can rise to $\pm 1.8934$ ppm for the farthest ISL that links $\textrm{LLO-1}$ and $\textrm{LLO-21}$ (inclination:$100^\circ$). It can be inferred from the subplots in Fig. \ref{fig:2} that, as the orbital inclination difference increases, the maximum relative velocity obtained from simulations increases as well, and hence the Doppler shift range becomes wider. This reveals the fact that further ISL in LSP requires higher Doppler tolerance.   

\begin{figure}[!t]
\centering
\subfloat[]{
	\label{subfig:4a}
	\includegraphics[width=.22\textwidth]{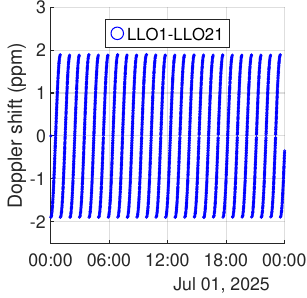}
	 }
\subfloat[]{
	\label{subfig:4b}
	\includegraphics[width=0.22\textwidth]{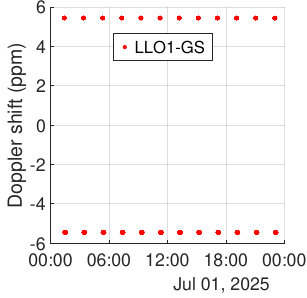}
	 }
\caption{(a) Time-varying ppm Doppler shifts between LLO-1 and LLO-21. \\
(b) Time-varying ppm Doppler shifts in intermittent links between LLO-1 and GS.}
\label{fig:4}
\end{figure}

Satellite-to-GS Doppler shift analyses are carried out between each LLO satellite and the GS located at Lunar South Pole ($0^{\circ}\mathrm{E}\,,\,90^{\circ}\mathrm{S}$). As shown in Fig. \ref{subfig:3b}, Doppler shifts are accumulated around minimum -5.4505 ppm and maximum 5.4509 ppm values rather than spreading. The intermittent links can be established with a duration of $\approx 10$ minutes during a period of $\approx 2$ hours. The Doppler shifts of $\textrm{LLO-1}$, which revolves 12 times in 1 day (i.e., simulation duration), are demonstrated in Fig. \ref{subfig:4b}. 

When inter-satellite and satellite-to-GS Doppler shifts are compared, it is clear that Doppler shifts are always higher in the second scenario, as demonstrated in Fig. \ref{fig:3} because the relative velocity between any LLO satellite and the GS at LSP is significantly high. It can be inferred that the communication system between the LLO satellite and the GS requires advanced methods to mitigate the time distortion caused by the Doppler frequency.

\begin{table*}[!h]
\centering
\caption{ Gaussian Mixture Model Parameters of Inter-satellite Links.}
\label{tab:my-GMM}
\resizebox{0.99\linewidth}{!}{%
\begin{tabular}{|l|lll|}
\toprule \toprule
\multicolumn{1}{|c|}{\multirow{2}{*}{\begin{tabular}[c]{@{}c@{}}\textbf{ISL}\end{tabular}}} & \multicolumn{3}{c|}{\textbf{Parameters}}                                                \\ \cline{2-4} 
\multicolumn{1}{|c|}{}                                                                   & \multicolumn{1}{c|}{$\pi$} & \multicolumn{1}{c|}{$\mu$} & \multicolumn{1}{c|}{$\sigma$} \\ \midrule
LLO-1 and -2  & \multicolumn{1}{l|}{(0.2288, 0.2229, 0.2913, 0.1309, 0.1260)} & \multicolumn{1}{l|}{(-0.066, 0.0660, -0.0011,-0.0918, 0.0917)} & \multicolumn{1}{l|}{(2.5677e-04, 2.6500e-04, 8.7808e-04, 8.9384e-06, 9.2748e-06)}  \\ \midrule

LLO-1 and -3  & \multicolumn{1}{l|}{(0.1358,0.1317,0.2734,0.2270,0.2321)} & \multicolumn{1}{l|}{(0.1825,-0.1835,-0.0080,-0.1333,0.1262)} &  \multicolumn{1}{l|}{(4.8835e-05, 3.6562e-05, 0.0031, 0.0010, 0.0012)} \\ \midrule

LLO-1 and -4  & \multicolumn{1}{l|}{(0.3195,0.2267,	0.1206,	0.2055,	0.1277)} & \multicolumn{1}{l|}{(0.0045,-0.2011,0.2759,0.2058,-0.2757)} &   \multicolumn{1}{l|}{(0.0093, 0.0023, 6.9993e-05, 0.0020, 7.3563e-05)} \\ \midrule

LLO-1 and -5  & \multicolumn{1}{l|}{(0.2353, 0.1375, 0.2621, 0.1358, 0.2292)} & \multicolumn{1}{l|}{(-0.2600,
0.3645,-0.0079,-0.3660,0.2526)} & \multicolumn{1}{l|}{(0.0044, 2.0280e-04, 0.0117, 1.6545e-04, 1.6545e-04)} \\ \midrule

LLO-1 and -6  & \multicolumn{1}{l|}{(0.1511,0.1351,	0.2542,	0.2305,	0.2291)} & \multicolumn{1}{l|}{(-0.4536, 0.4561, -0.0005, -0.3113, 0.3177)} &  \multicolumn{1}{l|}{(3.8385e-04, 2.9820e-04, 0.0183, 0.0075, 0.0075)} \\ \midrule 

LLO-1 and -7  & \multicolumn{1}{l|}{(0.1305,	0.1351,	0.2314,	0.2251,	0.2779)} & \multicolumn{1}{l|}{(-0.5502,
0.5472, 0.3799, -0.4014, -0.0248)} &  \multicolumn{1}{l|}{(3.2068e-04, 4.2692e-04, 0.0108, 0.0088, 0.0289)} \\ \midrule

LLO-1 and -8  & \multicolumn{1}{l|}{(0.1414, 0.2261,	0.1321,	0.2343,	0.2661)} & \multicolumn{1}{l|}{(0.6360,
-0.4656, -0.6411, 0.4322, -0.0353)} &  \multicolumn{1}{l|}{(6.8526e-04, 0.0122, 4.5846e-04, 0.0157, 0.0367)} \\ \midrule

LLO-1 and -9  & \multicolumn{1}{l|}{(0.2390, 0.1289, 0.2138,	0.1324,	0.2859)} & \multicolumn{1}{l|}{(-0.5200,
0.7319, 0.5312, -0.7324, 0.0134)} &  \multicolumn{1}{l|}{(0.0181, 6.3008e-04, 0.0157, 6.0630e-04, 0.0544)} \\ \midrule

LLO-1 and -10 & \multicolumn{1}{l|}{(0.1274, 0.2221, 0.3015,	0.2220,	0.1270)} & \multicolumn{1}{l|}{(-0.8260, 0.5914, -0.0227,-0.6083,0.8236)} &  \multicolumn{1}{l|}{(6.5393e-04, 0.0215, 0.0750, 0.0192, 7.6720e-04)}\\ \midrule

LLO-1 and -11 & \multicolumn{1}{l|}{(0.2601, 0.1478,	0.1270, 0.2507,	0.2145)} & \multicolumn{1}{l|}{(0.0573,
-0.9076, 0.9152, -0.6081, 0.6682)} &\multicolumn{1}{l|}{(0.0726, 0.0014, 9.3444e-04, 0.0340, 0.0240)} \\ \midrule

LLO-1 and -12 & \multicolumn{1}{l|}{(0.2215, 0.1459, 0.2803,	0.2261,	0.1261)} & \multicolumn{1}{l|}{(0.7263,
-0.9993, 0.01085, -0.6993, 1.0068)} &\multicolumn{1}{l|}{(0.0316, 0.0016, 0.1015, 0.0339, 0.0011)}  \\ \midrule

LLO-1 and -13 & \multicolumn{1}{l|}{(0.1843, 0.1519,	0.3049	0.2452,	0.1138)} & \multicolumn{1}{l|} {(0.8702,
-1.0849, -0.6421, 0.2468, 1.1055)} &\multicolumn{1}{l|}{(0.0220, 0.0023, 0.0755, 0.0903, 8.6624e-04)}  \\ \midrule

LLO-1 and -14  & \multicolumn{1}{l|}{(0.2567,	0.1315,	0.1442,	0.2449,	0.2226)} &\multicolumn{1}{l|}{(0.7612,
-1.1891,1.1773,-0.1203,-0.8717)} & \multicolumn{1}{l|}{(0.0655, 0.0015, 0.0026, 0.1072, 0.0395)} \\ \midrule

LLO-1 and -15 & \multicolumn{1}{l|}{(0.1432,	0.2213,	0.2722,	0.1285,	0.2348)} & \multicolumn{1}{l|}{(1.2681,
-0.9459,-0.0934,-1.2826,0.8548)} &\multicolumn{1}{l|}{(0.0029, 0.0454, 0.1517, 0.0016, 0.0641)}  \\ \midrule

LLO-1 and -16 & \multicolumn{1}{l|}{(0.3199, 0.2319,	0.2006,	0.1295,	0.1180)} & \multicolumn{1}{l|}{(0.0541,
-0.9894, 1.0403, -1.3731, 1.3771)} & \multicolumn{1}{l|}{(0.2309, 0.0599,   0.0460, 0.0019, 0.0016)} \\ \midrule

LLO-1 and -17 & \multicolumn{1}{l|}{(0.2580,	0.1372,	0.2046,	0.1230,	0.2772)} & \multicolumn{1}{l|}{(-0.9922,
-1.4582, 1.0939, 1.4651, 0.1154)} &\multicolumn{1}{l|}{(0.0867, 0.0028, 0.0544, 0.0021, 0.2044)}  \\ \midrule

LLO-1 and -18 & \multicolumn{1}{l|}{(0.2429,0.1283,	0.1388,	0.2694,	0.2206)} & \multicolumn{1}{l|}{(-1.0773,
1.5518, -1.5481, 0.0364, 1.1172)} &\multicolumn{1}{l|}{(0.0866,0.0028,0.0032,0.2201,0.0742)}  \\ \midrule

LLO-1 and -19 & \multicolumn{1}{l|}{(0.2998,	0.1339,	0.2114,	0.1231,	0.2317)} & \multicolumn{1}{l|}{(-0.1266,
1.6376, -1.2466, -1.6520, 1.1368)} & \multicolumn{1}{l|}{(0.2924, 0.0037, 0.0667, 0.0023, 0.0980)} \\ \midrule

LLO-1 and -20 & \multicolumn{1}{l|}{(0.1371, 0.1933,	0.2539,	0.2996,	0.1161)} & \multicolumn{1}{l|}{(-1.7305,
1.3404, -1.1831, 0.1759, 1.7449)} & \multicolumn{1}{l|}{(0.0039,0.0659, 0.1198, 0.3282, 0.0024)} \\ \midrule

LLO-1 and -21 & \multicolumn{1}{l|}{(0.3132, 0.1291,	0.1222,	0.2251,	0.2103)} & \multicolumn{1}{l|}{(-0.1120,
1.8237,-1.8351,1.2950,-1.3867)} & \multicolumn{1}{l|}{(0.3902, 0.0040, 0.0027, 0.1110, 0.0820)} \\ \midrule

ALL & \multicolumn{1}{l|}{(0.1426,	0.1213,	0.2435,	0.2207,	0.2718)} & \multicolumn{1}{l|}{(-1.2170, 1.2571, 0.2425,
-0.1595, -0.0575)} & \multicolumn{1}{l|}{(0.1200, 0.1085, 0.2954, 0.3348, 0.1636)} \\ \bottomrule \bottomrule
\end{tabular}%
}
\vspace{-0.35cm}
\end{table*}

The error metric results in terms of KL divergence and WMRD are exhibited in Table~\ref{tab:my-Error}. KL divergence always takes a non-negative value, and it has a range of $[0,\infty)$. Hence, the lower its value, the higher the resemblance of probability distributions. The maximum and minimum KL divergence values are 2.2963 and 0.0111, respectively; thus, the latter one is more identical to the corresponding GMM fitting curve than the former one. On the other hand, a zero WMRD means that no average relative difference; thus, the target value of WMRD is as close to zero as possible. The minimum and maximum values of WMRD are 0.5794 and 0.6906, respectively, which are fairly close to zero.  

The Doppler shift variations as a function of time are demonstrated in Fig. \ref{subfig:4a} during the simulation duration. When $\textrm{LLO-1}$ and $\textrm{LLO-21}$ move toward (away) each other $f_{20,d}>0$ ($f_{20,d}<0$). Doppler shift converges to zero when $\textrm{LLO-1}$ and $\textrm{LLO-21}$ have the same velocity, which leads to zero relative velocity. For instance, $f_{20,d}=0.00778$ at time 02:27:20 and the position and velocity vectors of $\textrm{LLO-1}$ and $\textrm{LLO-21}$ are [-345.95, -87.34, 1802.6] km and [292.33, -101.68, 1811.2] km, and [-0.038, -1.629, -0.085] km/s (speed: 1.6321 km/s) and [-0.039, -1.631, -0.086] km/s  (speed: 1.6340 km/s). Hence, the relative velocity vector, relative speed, and Doppler shift in ppm are computed as [-0.0010, -0.0020, -0.0010] km/s, 0.0023 km/s and 0.0078, respectively.

Table \ref{tab:my-GMM} presents the GMM with $K=5$  parameters; which are $\pi_{n,k}$ is the mixture coefficient representing the probability of $k$-th component, $\mu_{n,k}$ is the mean value of the $k$-th component, and $\sigma_{n,k}$ denotes the standard deviation value of the $k$-th component for the $n$-th ISL.

\section{Conclusion}
In this study, the Doppler shifts in realistic Cislunar inter-satellite and satellite-to-ground station scenarios are analyzed and statistically modelled. An ML estimator for both single ISLs and a more generalized ISL is derived for Doppler shift parameter estimation. Gaussian Mixture Models with $K=5$  components are used for curve fitting of these statistical Doppler shift models. Parts per million Doppler shifts allow us to model independently of a specific carrier frequency, which changes the Doppler frequency significantly. Furthermore, goodness-of-fit is evaluated by using Kullback–Leibler divergence and the weighted mean relative difference error metrics. 

The outcomes of the ISL simulations show that the higher the difference in inclination of the low lunar orbits, the larger the Doppler shift becomes due to the increasing relative velocity changes. Regarding the satellite-to-ground station scenario, the Doppler shift is concentrated in extreme values, as the relative velocity does not deviate significantly, unlike in the former case. This leads to the fact that the latter communication link requires more advanced systems, which can compensate for the distortions caused by the Doppler shifts. The Kullback–Leibler divergence and the weighted mean relative difference error metrics yield values near zero, and these results show that the proposed GMM with $K=5$ components establish good fit over the Doppler shift data in ppm.  

\section*{Acknowledgment}
 This work was supported in part by the Tier-1 Canada Research Chair program.

\bibliographystyle{IEEEtran}
\bibliography{bibliography}

\end{document}